\documentclass[aps,prd,preprint,showpacs,floats,nofootinbib]{revtex4-1}
\usepackage[dvips]{graphicx}
\usepackage{bm}
\usepackage{euscript,amsmath}

\begin{document}
\title{WIMPless dark matter and the excess gamma rays from the Galactic center }
\author {Guohuai Zhu}
\email[E-mail address: ]{zhugh@zju.edu.cn}
\affiliation{Zhejiang Institute of Modern Physics, Department of Physics, \\
 Zhejiang University, Hangzhou, Zhejiang 310027, P.R. China}

\date{\today}
\begin{abstract}
In this paper we discuss the excess gamma rays from the Galactic center, the WMAP haze and the CoGeNT and DAMA results in WIMPless models. At the same time we also investigate the low energy constraints from the anomalous magnetic moment of leptons and from some lepton flavor violating decays. It is found that, for scalar or vector WIMPless dark matter, neither the WMAP haze nor the CoGeNT and DAMA observations could be explained simultaneously with the excess gamma rays from the Galactic center.
As to fermion WIMPless dark matter, it is only marginally possible to accommodate the CoGeNT and DAMA results with the excess gamma rays from the Galactic center with vector connector fields. On the other hand, only scalar connector fields could interpret the WMAP haze concerning the constraints of anomalous magnetic moment of leptons. Furthermore, if there is only one connector field for all the charged leptons, some lepton flavor violating decays could happen with too large branching ratios severely violating the experimental bounds .
\end{abstract}
\pacs{12.90.+b,95.35.+d}
\maketitle
\section{Introduction}
There are lots of gravitational evidences suggesting the dominance of dark matter (DM) over baryonic visible matter in our universe,
though little is known about its non-gravitational interaction. If DM can annihilate or decay into the standard model (SM)
particles, gamma rays will be generated either directly or by the final state radiation, inverse Compton scattering and bremsstrahlung of the final
charged particles. As these gamma rays depend on the DM density as $\rho^{2}$ for annihilations and $\rho$ for decays, they might be detectable at locations with very high density of dark matter, such as the Galactic center (GC).

Recently Hooper and Goodenough \cite{Hooper:2010mq} investigated the gamma rays from the GC using the first two years of data from the  Fermi Gamma Ray Space Telescope. They modeled the backgrounds and found that the morphology and spectrum of the gamma rays between $1.25^\circ$ and $10^\circ$ from the GC can be well understood. However excess emission of gamma rays seems to be present within $1.25^\circ$ degree of the GC. This may well indicate that we have to improve our understanding on the astrophysical background.
Nevertheless, this additional component could also be interpreted by the DM annihilation into tau leptons, with a mass around $8$ GeV and the annihilation cross section $\langle \sigma v \rangle \sim 10^{-26}\mbox{cm}^3/$s.

There are also anomalous microwave emission
from the inner Galaxy in the WMAP data \cite{Finkbeiner:2003im,Dobler:2007wv} which is known as the "WMAP haze".
It was further observed by Hooper and Linden \cite{Hooper:2010im} that, if the DM annihilates equally into $e^+ e^-$, $\mu^+ \mu^-$ and
$\tau^+ \tau^-$, the WMAP haze can be simultaneously accounted for by the synchrotron radiation emitted from the annihilation final state of electrons and positrons. Notice that WMAP haze could also be explained by astrophysical processes \cite{Crocker:2010qn}.

Coincidentally, the annual modulation effect measured by the DAMA/LIBRA
collaboration \cite{Bernabei:2010mq} together with the recent results about excesses of
events (over the backgrounds) reported by CDMS-II \cite{Ahmed:2009zw} and CoGeNT \cite{Aalseth:2010vx},
if interpreted as elastically scattering DM, favor the DM mass about 5
- 10 GeV \cite{Bottino:2009km,Kopp:2009qt,Fitzpatrick:2010em,Chang:2010yk,Foot:2010rj}. The null results from Xenon100 \cite{Aprile:2010um} and CDMS-II \cite{Ahmed:2010wy} are in tension with the previous data. Nevertheless, concerning the
uncertainties of the DM velocity distribution \cite{Lisanti:2010qx} and of the relative scintillation efficiency of the liquid xenon at low recoil energy \cite{Savage:2010tg,Collar:2010nx,Collar:2010ht}, it is still possible to interpret the CoGeNT and DAMA/LIBRA observations as light DM with mass around $7$ GeV \cite{Hooper:2010uy}, and at the same time, consistent with the null results of Xenon100 and CDMS-II.

Such a light DM is beyond the conventional paradigm of weakly-interacting massive particle (WIMP) with the mass around $100$ GeV to several TeV, though it is certainly possible to realize this scenario in SUSY models, e.g. in an effective minimal supersymmetric extension of the Standard Model \cite{Bottino:2003iu,Bottino:2003cz,Bottino:2009km,Fornengo:2010mk,Scopel:2011qt}.
Interestingly, in the WIMPless models \cite{Feng:2008ya} the DM particle naturally has the correct thermal relic density with a wide range of mass varying from TeV down to subGeV \cite{Feng:2008mu}. The key observation in \cite{Feng:2008ya} is that, the thermal relic density of a stable particle is proportional to $m^2/g^4$, with $m$ and $g$ the typical mass and coupling entering the annihilation cross section.
The correct thermal relic density can be easily realized for the DM particle X in the hidden sector if
\begin{equation}
 \frac{m_X}{g_X^2} \sim \frac{m_{weak}}{g_{weak}^2}~,
\end{equation}
with the weak scale $m_{weak}\simeq 100$ GeV and $g_{weak}\simeq 0.65$. 
Such relation can be satisfied for example in the gauge-mediated supersymmetry breaking models if there is only one supersymmetry(SUSY) breaking sector, so that the SUSY-breaking soft scales in both the visible sector and the hidden sector are generated by the gauge interactions with the
same SUSY-breaking sector \footnote{There exists many light DM scenarios, of which dark atoms of DM is an exotic example \cite{Khlopov:2010pq}}.

In the following, we will then study the possible implications of the above experimental results for the WIMPless models.
Notice that we shall not attempt to explore the full parameter space of the WIMPless models. Instead, as illustration only benchmark models are examined in this paper. For example, following Ref. \cite{Hooper:2010im}, we only consider the WIMPless DM annihilating democratically into $e^+ e^-$, $\mu^+ \mu^-$ and $\tau^+ \tau^-$ to interpret the WMAP haze. But it is definitely possible to explain the WMAP haze with other choices of parameters provided the DM can significantly annihilate into $e^+ e^-$. Nevertheless, our main conclusions should remain qualitatively unchanged for general WIMPless models.

Phenomenologically, the WIMPless DM has been used to explain the DAMA signal \cite{Feng:2008dz}. The indirect detection of fermion WIMPless DM at the neutrino telescopes IceCube and DeepCore has been discussed in \cite{Barger:2010ng}. The low energy constraints on WIMPless DM can be found in \cite{McKeen:2009rm,McKeen:2009ny}.  It is also worth noting that the particle physics implications for the gamma ray excess from the GC together with CoGeNT and DAMA/LIBRA has been studied in \cite{Buckley:2010ve}, though in a model independent way and without the inclusion of the WMAP haze.

This paper is organized as follows. In the next section, we adopt scalar and vector WIMPless models to discuss the excess gamma rays from the GC, the WMAP haze and the CoGeNT and DAMA results. We also investigate the low energy constraints from the anomalous magnetic moment of leptons and from some lepton flavor violating decays.  We then extend the discussions to fermion WIMPless models in the
third section. Finally we conclude with a summary in section IV.

\section{The scalar and vector WIMPless DM}

In WIMPless models, to couple the DM particle X to the visible sector, Yukawa like interactions including a connector sector can be introduced.
To be concrete, let's first consider the case of scalar DM which was exploited in \cite{Feng:2008dz} to investigate the DAMA annual modulation signals. The interactions can be written as
\begin{align}\label{Eq:Yukawa}
        {\cal L}_{int}= \lambda_{f_L} X \bar{Y}^f_R f_L + \lambda_{f_R} X \bar{Y}^f_L f_R + h.c.~,
\end{align}
where $f_L$ and $f_R$ denote the left-handed and right-handed Standard Model (SM) fermions, $Y^f$ are the connector fields which have both hidden and SM charges. Notice that $Y^f$ behave just like exotic fourth generation quarks and leptons (or squarks and sleptons). The direct searches at Tevatron constrain the masses of the exotic fourth generation quarks to be $Y^b >330$ GeV \cite{Alwall:2010jc}, while the searches at LEP constrain the sleptons to be heavier than $\sim 100$ GeV \cite{LEPSUSY}. Therefore the connector fields Y must be much heavier than
the DM particle X which is about $8$ GeV, and in the following calculations we will take the limit of large $m_Y$. $X$ can also be a vector field,
correspondingly the interactions is
\begin{align}\label{Eq:Yukawa-vector}
        {\cal L}_{int}= \lambda_{f_L} X^\mu \bar{Y}^f_L \gamma_\mu f_L + \lambda_{f_R} X^\mu \bar{Y}^f_R \gamma_\mu f_R + h.c.~
\end{align}

With the above lagrangians, the DM particles can annihilate into the SM fermions as shown in Fig. 1a,b. Then
\begin{align}
\langle \sigma v \rangle &\simeq \frac{\lambda_{f_L}^2 \lambda_{f_R}^2}{2\pi m_{Y^f}^2} \left (  1-m_f^2/m_X^2 \right )^{3/2}
 \hspace*{1cm} \textrm{(for scalar X)} \\
\langle \sigma v \rangle &\simeq \frac{\lambda_{f_L}^2 \lambda_{f_R}^2}{6\pi m_{Y^f}^2} \left (  1-m_f^2/m_X^2 \right )^{3/2}
 \hspace*{1cm} \textrm{(for vector $X^\mu$)}
\end{align}
It is shown in \cite{Hooper:2010mq} that the excess gamma rays from the GC can be explained by the DM annihilation primarily into tau lepton pairs, with the annihilation cross section $\langle \sigma v \rangle$ in the range of  $3.3 \times 10^{-27} \mbox{cm}^3/$s to $1.5 \times 10^{-26}\mbox{cm}^3/$s. Taking $m_X=8$ GeV, This requires
\begin{align}
     \frac{\lambda_{\tau_L} \lambda_{\tau_R}}{m_{Y^\tau}} \simeq \left \{ \begin{array}{ll}
     (4.3-9.2) \times 10^{-5} \mbox{~GeV}^{-1} & \textrm{(for scalar X)} \\
     (7.5-16.0) \times 10^{-5} \mbox{~GeV}^{-1} & \textrm{(for vector $X^\mu$)} \end{array} \right.
\end{align}
It was also pointed out in \cite{Hooper:2010mq} that the DM X may annihilate into ${\bar b} b$ or ${\bar c}c$ final states up to $20$\% of the time without spoiling the DM interpretation of the excess gamma rays from the GC. This means
\begin{align}\label{Eq:GCconstraint}
 \frac{\lambda_{b_L} \lambda_{b_R}}{m_{Y^b}}   \leq 0.65 \frac{\lambda_{\tau_L} \lambda_{\tau_R}}{m_{Y^\tau}}
 \simeq \left \{ \begin{array}{ll}(2.8 - 6.0) \times 10^{-5} \mbox{~GeV}^{-1}  & \textrm{(for scalar X)} \\
 (4.9 - 10.4) \times 10^{-5} \mbox{~GeV}^{-1} & \textrm{(for vector $X^\mu$)} \end{array} \right.
\end{align}

\begin{figure}
\includegraphics[scale=1.0]{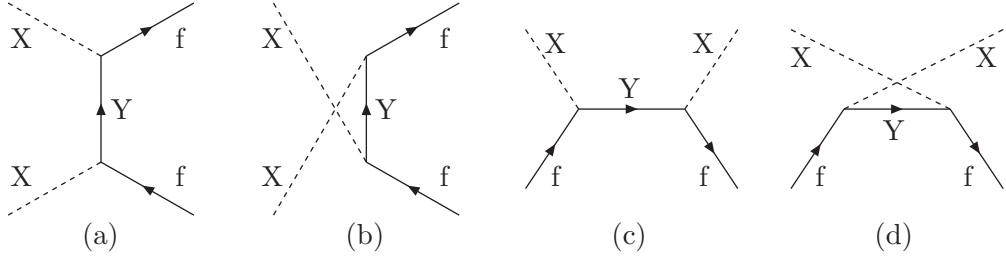} \caption{\label{fig:1} The Feynman diagrams for (a), (b): the scalar DM annihilation process $XX \to f\bar{f}$ and (c),(d): the scalar DM-bottom quark scattering. For the case of vector DM, the diagrams are the same except for the external scalar bosons to be  replaced by the vector bosons}
\end{figure}

The interaction terms in Eqs.(\ref{Eq:Yukawa},\ref{Eq:Yukawa-vector}) can also induce spin-independent X-nucleon scattering as shown in Fig. 1c,d.  To escape flavor changing neutral current (FCNC) constraints, we will assume following \cite{Feng:2008dz} that only $\lambda_b$ and $\lambda_t$ are significant for X-Y-quark couplings. The corresponding direct detection cross section \cite{Feng:2008dz} is the same for both scalar X and vector $X^\mu$, as
\begin{align}
  \sigma_{SI}^{p,n} \simeq \frac{m_{p,n}^2}{4\pi (m_X+m_{p,n})^2} \frac{4\lambda_{b_L}^2 \lambda_{b_R}^2}{m_{Y^b}^2}
     \left ( \frac{2 m_{p,n}f_g^{p,n}}{27 m_b} \right )^2~,
\end{align}
where $f_g^{p,n}=1-f_u^{p,n}-f_d^{p,n}-f_s^{p,n}\simeq (0.43-0.74)$ \cite{Bottino:2001dj,Ellis:2005mb} with the uncertainties coming mainly from
$f_s^{p,n}$. The CoGeNT and DAMA/LIBRA results, if interpreted
as DM elastic scattering, require $\sigma_{SI}^{p,n}\sim (1-2) \times 10^{-40} \mbox{cm}^2$
\cite{Kopp:2009qt,Fitzpatrick:2010em,Chang:2010yk,Foot:2010rj,Hooper:2010uy}\footnote{Notice that this range of cross section is obtained by assuming that the CoGENT excess is actually significant of a real signal. The DAMA/LIBRA results, take alone, are compatible with a much larger range of values for the elastic cross section.}. This implies the following constraint
\begin{align}
  \frac{\lambda_{b_L} \lambda_{b_R}}{m_{Y^b}} \sim (6.9-16.8) \times 10^{-4} \mbox{GeV}^{-1} ~,
\end{align}
which completely contradicts Eq.(\ref{Eq:GCconstraint}). Therefore it seems to be difficult, if not impossible, for the scalar or vector WIMPless DM to account for the gamma ray excess from the GC together with the direct detection results of CoGeNT and DAMA.

Furthermore, even if we forget about the direct detection signals, there are stringent constraints, if considering the WMAP haze, from anomalous magnetic moment of leptons and some lepton flavor violating decays. As analyzed in \cite{Hooper:2010im}, the WMAP haze could be accounted for
by the DM annihilation democratically into $e^+ e^-$, $\mu^+ \mu^-$ and $\tau^+ \tau^-$, which means
\begin{align}\label{Eq:Hazeconstraint}
\frac{\lambda_{e_L} \lambda_{e_R}}{m_{Y^e}} \simeq \frac{\lambda_{\mu_L} \lambda_{\mu_R}}{m_{Y^\mu}} \simeq
\frac{\lambda_{\tau_L} \lambda_{\tau_R}}{m_{Y^\tau}} \simeq \left \{ \begin{array}{ll}
     (4.3-9.2) \times 10^{-5} \mbox{~GeV}^{-1} & \textrm{(for scalar X)} \\
     (7.5-16.0) \times 10^{-5} \mbox{~GeV}^{-1} & \textrm{(for vector $X^\mu$)} \end{array} \right.
\end{align}
But as shown in Fig 2(a), the $XYl$ couplings also contribute to the anomalous magnetic moment of leptons. For the scalar case, it is found to be
\footnote{our expression is different from that derived in \cite{McKeen:2009ny}, which is probably due to that the author of
\cite{McKeen:2009ny} did not notice that the hermitian conjugate of $\bar{Y}^f(1-\gamma_5)f$ should be $\bar{f}(1+\gamma_5)Y^f$.}
\begin{align}
\Delta a_l = \frac{1}{16\pi^2} \int_0^1~dx~(1-x)^2\frac{2\lambda_{l_L}\lambda_{l_R}m_{Y^l}m_l+x(\lambda_{l_L}^2+\lambda_{l_R}^2)m_l^2}
{x m_X^2 + (1-x)m_{Y^l}^2} \simeq \frac{\lambda_{l_L}\lambda_{l_R}m_l}{16\pi^2 m_{Y^l}}~,
\end{align}
while for the vector case, it is
\begin{align}\label{vector-amu}
\Delta a_l & \simeq  \frac{1}{16\pi^2} \int_0^1~dx \frac{8x(1-x)\lambda_{l_L}\lambda_{l_R}m_{Y^l}m_l
-(1-x)^3(\lambda_{l_L}^2+\lambda_{l_R}^2)m_{Y^l}^2 m_l^2/m_X^2}{x m_X^2 + (1-x)m_{Y^l}^2} \nonumber \\
& \simeq \frac{\lambda_{l_L}\lambda_{l_R}}{4\pi^2}\frac{m_l}{m_{Y^l}}-\frac{\lambda_{l_L}^2+\lambda_{l_R}^2}{48\pi^2}\frac{m_l^2}{m_X^2}
\end{align}

There is well known $3\sigma$ deviation for the muon anomalous magnetic moment between the SM expectation and experimental measurement as \cite{Nakamura:2010zzi}
\begin{align}
\Delta a_\mu=a_\mu^{exp}-a_\mu^{SM} = (2.55 \pm 0.80) \times 10^{-9}~.
\end{align}
For the scalar case, this limits
\begin{align}
\frac{\lambda_{\mu_L}\lambda_{\mu_R}}{m_{Y^\mu}} < 3.8 \times 10^{-6} \left ( \frac{\Delta a_\mu}{2.55 \times 10^{-9}}\right ) \mbox{GeV}^{-1}
\end{align}
which strongly violates Eq.(\ref{Eq:Hazeconstraint}). For the vector case, assuming $\lambda_{1_L}=\lambda_{1_R}$, the second term of Eq.(\ref{vector-amu}) is about one order smaller than the first term, and we obtain
\begin{align}
\frac{\lambda_{\mu_L}\lambda_{\mu_R}}{m_{Y^\mu}} < 1.1 \times 10^{-6} \left ( \frac{\Delta a_\mu}{2.55 \times 10^{-9}}\right ) \mbox{GeV}^{-1}
\end{align}
which again severely violates Eq.(\ref{Eq:Hazeconstraint}).

For the electron, the experimental data gives $a_e^{exp} = (1159652180.7
\pm 0.3) \times 10^{-12}$ \cite{Nakamura:2010zzi} which agrees well with the SM calculation
$a_e^{SM}=(1159652182.8 \pm 7.7) \times 10^{-12}$\cite{Aoyama:2007mn}.
We will assume $\Delta a_e < 10 \times 10^{-12}$ which is slightly larger than the theoretical error,
to get a restriction on the corresponding $XYe$ couplings
\begin{align}
\frac{\lambda_{e_L}\lambda_{e_R}}{m_{Y^e}} &<  3.1 \times 10^{-6} \left ( \frac{\Delta a_e}{10 \times 10^{-12}}\right ) \mbox{GeV}^{-1}
 \hspace*{1cm} \textrm{(for scalar X)} \\
\frac{\lambda_{e_L}\lambda_{e_R}}{m_{Y^e}} &<  7.7 \times 10^{-7} \left ( \frac{\Delta a_e}{10 \times 10^{-12}}\right ) \mbox{GeV}^{-1}
\hspace*{1cm} \textrm{(for vector $X^\mu$)}
\end{align}
which also strongly violates Eq.(\ref{Eq:Hazeconstraint}).

If one connector field $Y^{lep}$ can couple to different generations of leptons, namely $Y^e=Y^\mu=Y^\tau=Y^{lep}$, the $XYl$ interactions could also induce the lepton flavor violating (LFV) $\tau \to \mu \gamma$, $e\gamma$ and $\mu \to e \gamma$ decays, as shown in Fig. 2(b). As the anomalous magnetic moment of leptons already exclude the possibility for scalar or vector WIMPless DM to explain the WMAP haze together with the excess gamma rays from the GC, we will not investigate further these LFV decays. But in the next section these LFV decays will turn out to be important in the case of fermion WIMPless DM.

\begin{figure}
\includegraphics[scale=1.2]{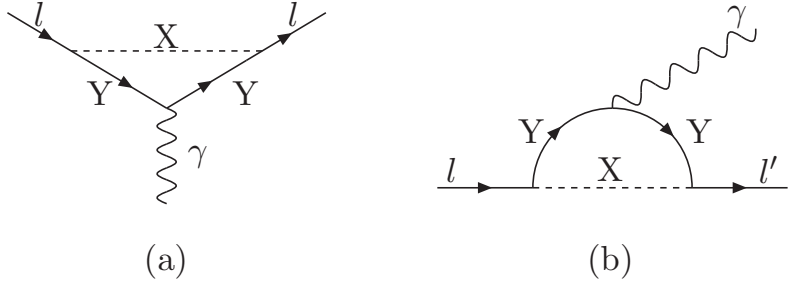} \caption{\label{fig:2} The $XYl$ interactions contribute to (a) the anomalous magnetic moment of leptons and
(b) the lepton flavor violating $\tau \to \mu \gamma$, $e\gamma$ and $\mu \to e \gamma$ decays with scalar X. For the case of vector $X^\mu$, the diagrams are the same except for the internal scalar line to be replaced by the vector line.}
\end{figure}

Therefore in scalar or vector WIMPless models, neither the WMAP haze nor the CoGeNT and DAMA observations could be explained simultaneously with the excess gamma rays from the GC.

\section{Fermion WIMPless DM}

If the DM particle $X$ is a fermion, the connector fields $Y$ should be either scalar or vector fields. For the scalar $Y$ case, the interactions can be written as
\begin{align}\label{Eq:fermion-Yukawa}
        {\cal L}_{int}= \lambda_{f_L} Y_{fL} \bar{X} f_L + \lambda_{f_R} Y_{fR} \bar{X} f_R + h.c.~,
\end{align}
where $Y_{fL}$ and $Y_{fR}$ are different scalar fields carrying different SM charges. For the vector $Y$ case, the corresponding interactions are
\begin{align}
 {\cal L}_{int}= \lambda_{f_L} Y^\nu_{fL} \bar{X} \gamma_v f_L + \lambda_{f_R} Y^\nu_{fR} \bar{X}\gamma_\nu f_R + h.c.~,
\end{align}
here $Y_{fL}$ and $Y_{fR}$ are different vector fields.

With the above lagrangians, the DM particles can annihilate into the SM fermions as shown in Fig. 3(a) with the cross section
\begin{align}
\langle \sigma v \rangle &\simeq \left ( \frac{\lambda_{f_L}^4}{m_{Y_{fL}}^4}+\frac{\lambda_{f_R}^4}{m_{Y_{fR}}^4}\right ) \frac{m_X^2}{32\pi}  \sqrt{1-m_f^2/m_X^2} \hspace*{1cm} \textrm{(for scalar Y)} \\
\langle \sigma v \rangle &\simeq \left ( \frac{\lambda_{f_L}^4}{m_{Y_{fL}}^4}+\frac{\lambda_{f_R}^4}{m_{Y_{fR}}^4}\right ) \frac{m_X^2}{8\pi}  \sqrt{1-m_f^2/m_X^2} \hspace*{1cm} \textrm{(for vector $Y^\nu$)}
\end{align}
Assuming for simplicity $\lambda_{f_L}/m_{Y_{fL}}=\lambda_{f_R}/m_{Y_{fR}}\equiv\lambda_{f}/m_{Y_{f}}$ and taking $m_X=8$ GeV,
the excess gamma rays from the GC restricts
\begin{align}\label{Eq:fermion-GC}
\frac{\lambda_{\tau}}{m_{Y_{\tau}}} = \left \{ \begin{array}{ll} (3.9-5.6)\times 10^{-3}~\mbox{GeV}^{-1} & \textrm{(for scalar Y)} \\
(2.7-4.0)\times 10^{-3}~\mbox{GeV}^{-1} & \textrm{(for vector $Y^\nu$)} \end{array} \right.
\end{align}
which corresponds to the DM annihilation into $\tau^+ \tau^-$ with the cross section $\langle \sigma v \rangle =(3.3-15) \times 10^{-27} \mbox{cm}^3/$s. Equivalently this means $m_{Y_{\tau}} \simeq \lambda_{\tau}(180-260)$ GeV for scalar Y and $m_{Y_{\tau}} \simeq \lambda_{\tau}(250-370)$ GeV for vector $Y^\nu$. As $Y_\ell$ behaves just like slepton, this does not contradict, if $\lambda$ is not much smaller than one, with the direct searches at LEP which constrain the sleptons to be heavier than $\sim 100$ GeV \cite{LEPSUSY}.

Considering direct detection experiments, the DM-nucleon scattering, as shown in Fig. 3(b), should be evaluated. In the large $m_Y$ limit and after a Fierz transformation, the DM-quark scattering can be written as effective operators
\begin{align}
 Q_{eff} &\simeq \frac{\lambda_{f}^2}{2m_{Y_f}^2}\left (\bar{f}\gamma^\mu P_R f\bar{X}\gamma_\mu P_L X+
                     \bar{f}\gamma^\mu P_L f\bar{X}\gamma_\mu P_R X \right ) \hspace*{1cm} \textrm{(for scalar Y)} \\
 Q^V_{eff} &\simeq\frac{\lambda_{f}^2}{m_{Y_f}^2}\left (\bar{f}\gamma^\mu P_L f\bar{X}\gamma_\mu P_L X +
            \bar{f}\gamma^\mu P_R f\bar{X}\gamma_\mu P_R X\right ) \hspace*{1cm} \textrm{(for vector $Y^\nu$)}
\end{align}
Notice that the spin-independent and velocity-independent contribution to DM-nucleon scattering comes only from the vector-vector interaction. Assuming $\lambda_u/m_{Y_u}=\lambda_d/m_{Y_d}\equiv\lambda_H/m_{Y_H}$, the DM-nucleon scattering cross section is
\begin{align}
\sigma_{SI}^{p,n}&\simeq \frac{m_X^2 m_{p,n}^2}{64\pi(m_X+m_{p,n})^2} \frac{9\lambda_H^4}{16m_{Y_H}^4} \hspace*{1cm} \textrm{(for scalar Y)} \\
\sigma_{SI}^{p,n}&\simeq \frac{m_X^2 m_{p,n}^2}{64\pi(m_X+m_{p,n})^2} \frac{9\lambda_H^4}{4m_{Y_H}^4} \hspace*{1cm} \textrm{(for vector $Y^\nu$)}
\end{align}
To obtain $\sigma_{SI}^{p,n}\sim (1-2) \times 10^{-40} \mbox{cm}^2$ suggested by the CoGeNT and DAMA results, one finds
\begin{align}
    \frac{\lambda_H}{m_{Y_H}} \sim \left \{ \begin{array}{ll} (3.4-4.0)\times 10^{-3}~\mbox{GeV}^{-1} & \textrm{(for scalar Y)} \\
    (2.4-2.8)\times 10^{-3}~\mbox{GeV}^{-1} & \textrm{(for vector $Y^\nu$)} \end{array} \right.
\end{align}
Equivalently this means $m_{Y_H} \sim \lambda_{H}(250-290)$ GeV for scalar Y and $m_{Y_H} \simeq \lambda_{\tau}(360-410)$ GeV for vector $Y^\nu$. As $Y_H$ behaves just like exotic fourth generation quark, the scalar Y case seems to be in disagreement, if $\lambda_H$ is smaller than one, with the Tevatron direct searches which limit the exotic fourth generation quark to be heavier than $330$ GeV \cite{Alwall:2010jc}. While for the vector $Y^\nu$ case, it is only  marginally consistent with the Tevatron limit.

\begin{figure}
\includegraphics[scale=1.0]{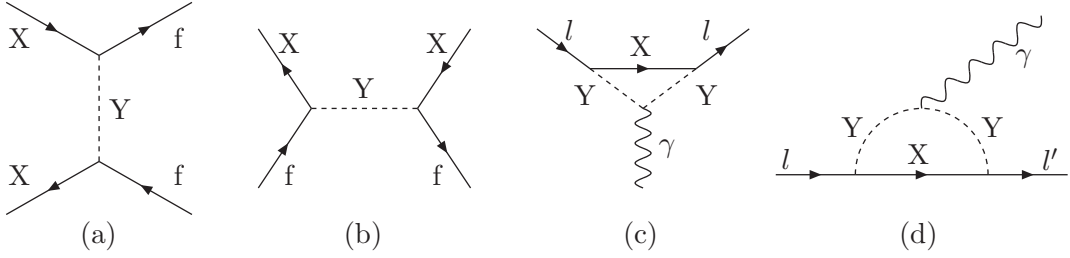} \caption{\label{fig:3} The Feynman diagrams for (a) the DM annihilation process $XX \to f\bar{f}$,
(b) the DM-quark scattering, (c) the anomalous magnetic moment of leptons and
(d) the lepton flavor violating $\ell \to \ell' \gamma$ decay with the scalar connector fields. For the vector $Y^\nu$ case, the diagrams are
the same except for all internal scalar lines to be replaced by the vector lines.}
\end{figure}

To explain the WMAP haze in the same framework, the DM candidate X should annihilate democratically into $e^+ e^-$, $\mu^+ \mu^-$ and $\tau^+ \tau^-$, which means
\begin{align}\label{Eq:fermion-haze}
\frac{\lambda_{e}}{m_{Y_{e}}}\simeq\frac{\lambda_{\mu}}{m_{Y_{\mu}}}\simeq\frac{\lambda_{\tau}}{m_{Y_{\tau}}} = \left \{ \begin{array}{ll}
(3.9-5.6)\times 10^{-3}~\mbox{GeV}^{-1} & \textrm{(for scalar Y)} \\
(2.7-4.0)\times 10^{-3}~\mbox{GeV}^{-1}  & \textrm{(for vector $Y^\nu$)} \end{array} \right.
\end{align}
The anomalous magnetic moment of leptons, as shown in Fig. 3(c), can be evaluated as
\begin{align}
\Delta a_l = -\frac{\lambda_{l_L}^2 m_l^2}{16\pi^2} \int_0^1~dx~\frac{x(1-x)^2}
{x m_X^2 + (1-x)m_{Y_{lL}}^2} +(L \rightarrow R) ~\simeq -\frac{\lambda_l^2 m_l^2}{48\pi^2 m_{Y_{l}}^2}~,
\end{align}
for the case of scalar connector fields. Similarly the contribution of vector connector fields to lepton's anomalous magnetic moment is
\begin{align}
\Delta a_l = -\frac{\lambda_{l_L}^2 m_l^2}{16\pi^2} \int_0^1~dx~\frac{(1-x)^2(3-2x)}
{x m_X^2 + (1-x)m_{Y_{lL}}^2} +(L \rightarrow R) ~\simeq -\frac{7\lambda_l^2 m_l^2}{48\pi^2 m_{Y_{l}}^2}~.
\end{align}
Notice that in both cases, the above contributions to lepton's anomalous magnetic moment are negative, which is opposite in sign to the $3\sigma$ deviation of the muon's anomalous magnetic moment $\Delta a_\mu^{exp}-\Delta a_\mu^{SM}=(2.55 \pm 0.85) \times 10^{-9}$. Therefore it is impossible for the fermion WIMPless models to account for the $3\sigma$ deviation of $a_\mu$. Instead the WIMPless contribution should be small enough, say $\vert \Delta a_\mu^{WIMPless} \vert < 0.85 \times 10^{-9}$, so that the situation does not become significantly worse.

Then one can obtain the following limits for the scalar Y case
\begin{align}
 \frac{\lambda_{\mu}}{m_{Y_{\mu}}} &< 6.0 \times 10^{-3} \left ( \frac{\Delta a_\mu}{0.85 \times 10^{-9}}\right )^{1/2} \mbox{GeV}^{-1}~, \nonumber \\
 \frac{\lambda_e}{m_{Y_e}} &< 0.13 \left ( \frac{\Delta a_e}{10 \times 10^{-12}}\right )^{1/2} \mbox{GeV}^{-1}~,
\end{align}
which are consistent with Eq.(\ref{Eq:fermion-haze}). However for the vector connector fields, the limits are
\begin{align}
 \frac{\lambda_{\mu}}{m_{Y_{\mu}}} &< 2.3 \times 10^{-3} \left ( \frac{\Delta a_\mu}{0.85 \times 10^{-9}}\right )^{1/2} \mbox{GeV}^{-1}~, \nonumber \\
 \frac{\lambda_e}{m_{Y_e}} &< 0.05 \left ( \frac{\Delta a_e}{10 \times 10^{-12}}\right )^{1/2} \mbox{GeV}^{-1}~,
\end{align}
where unfortunately the muon's $g-2$ leads to a constraint too stringent to be satisfied by Eq.(\ref{Eq:fermion-haze}).

Furthermore, if one connector field $Y^{lep}$ can couple to any charged leptons, namely $Y_e=Y_\mu=Y_\tau=Y^{lep}$, it could also induce the lepton flavor violating $\tau \to \mu \gamma$, $e\gamma$ and $\mu \to e \gamma$ decays, as shown in Fig. 3(d). Notice that the photon can
also be attached to external lepton lines, but those diagrams are gauge-dependent and should be canceled by the gauge-dependent part of
Fig. 3(d). Then taking the final state lepton to be massless and assuming for simplicity $\lambda_{l_L}=\lambda_{l_R}\equiv \lambda_l$,
the decay width is found to be
\begin{align}
\Gamma(\ell \to \ell' \gamma)&\simeq \frac{\alpha m_l^5}{1024\pi^4}\frac{\lambda_l^2 \lambda_{l'}^2}{72 m_{Y^{lep}}^4} \hspace*{1cm} \textrm{(for scalar Y)} \\
\Gamma(\ell \to \ell' \gamma)&\simeq \frac{\alpha m_l^5}{1024\pi^4}\frac{49\lambda_l^2 \lambda_{l'}^2}{72 m_{Y^{lep}}^4} \hspace*{1cm} \textrm{(for vector $Y^\nu$)}
\end{align}
Experimentally ${\cal B}(\mu \to e \gamma)<1.2\times 10^{-11}$, ${\cal B}(\tau \to e \gamma)<3.3\times 10^{-8}$ and
${\cal B}(\tau \to \mu \gamma)<4.4\times 10^{-8}$ \cite{Nakamura:2010zzi} lead to the constraints
\begin{align}
\frac{\sqrt{\lambda_\mu\lambda_e}}{m_{Y^{lep}}} &< 1.3 \times 10^{-4} (4.8 \times 10^{-5})~\mbox{GeV}^{-1}~,  \nonumber \\
\frac{\sqrt{\lambda_\tau\lambda_e}}{m_{Y^{lep}}} &< 1.4 \times 10^{-3}(5.4 \times 10^{-4}) ~\mbox{GeV}^{-1}~, \\
\frac{\sqrt{\lambda_\tau\lambda_\mu}}{m_{Y^{lep}}}  &< 1.5 \times 10^{-3} (5.8 \times 10^{-4})~\mbox{GeV}^{-1}~. \nonumber
\end{align}
for the case of scalar (vector) connector fields.
Unfortunately all the above limits are in clear contradiction with the requirement of the WMAP haze of Eq.(\ref{Eq:fermion-haze}).

\section{Summary}
In this paper we discuss in WIMPless models the possibility to interpret the excess gamma rays from the Galactic center, the WMAP haze and the CoGeNT and DAMA results. At the same time some low energy constraints must be satisfied, such as the anomalous magnetic moment of leptons and lepton flavor violating $\ell' \to \ell \gamma$ decays. Notice that a connector sector Y is introduced to couple the DM candidate X to the SM particles via Yukawa-like interactions.

As shown in \cite{Hooper:2010mq}, the excess gamma rays from the GC implies that the DM particles should annihilate dominantly into lepton pairs with  ${\bar b} b$ or ${\bar c}c$ final states less than $20$\% of the time. For scalar or vector WIMPless DM, this limits the coupling strength of $XYb$ to be too small to account for the CoGeNT and DAMA observations. To interpret the WMAP haze, the $XY\ell$ coupling is required to be roughly universal for $e$, $\mu$ and $\tau$ leptons, which will lead to too large contribution to anomalous magnetic moment of electron and muon.

As to fermion WIMPless DM with scalar or vector connector sector, $\lambda/m_Y \sim $ a few $\times 10^{-3}~\mbox{GeV}^{-1}$ could accommodate the excess gamma rays from the GC and the CoGeNT and DAMA results. This corresponds to a scalar connector particle $Y_H$ less than $300$ GeV or a vector $Y_H$ less than about $400$ GeV. As $Y_H$ couples to quarks, it behaves like an exotic fourth generation quark, which has been restricted by the Tevatron to be heavier than $330$ GeV \cite{Alwall:2010jc}. Therefore the scalar Y case seems to be disfavored, while the vector $Y^\nu$ case is only  marginally consistent with the Tevatron limit. Instead, to interpret the WMAP haze in the same framework, the constraints of anomalous magnetic moment of leptons can only be satisfied for the scalar connector fields.
Furthermore, if there is only one connector field for charged leptons, the lepton flavor violating $\ell \to \ell' \gamma$ decays could happen in both cases with too large branching ratios severely violating the experimental bounds .

Therefore it is difficult, if not impossible, for the scalar or vector WIMPless DM to interpret the excess gamma rays from the GC together with either the WMAP haze or the direct detection experiments CoGeNT and DAMA. While for fermion WIMPless DM, it may be possible to accommodate the excess gamma rays from the GC and the CoGENT and DAMA results with vector connector fields, though it is just marginally consistent with the Tevatron limit. On the contrary, only scalar connector fields could explain the WMAP haze under the constraints of anomalous magnetic moment of the leptons. In addition, if one connector field can couple to any charged leptons, the lepton flavor violating $\ell \to \ell' \gamma$ decays could happen with the branching ratios surpassing the current experimental bounds by (at most) even several orders of magnitude.

Quantitatively, the above conclusions are only valid in the WIMPless models. But going beyond this specific model, one may at least learn that the excess gamma rays from the Galactic center, if interpreted as DM annihilation, could impose strong constraint on the DM couplings to quarks.
Then one should check, under this constraint, the interpretation of the CoGENT and DAMA results in terms of elastic DM-nucleon scattering. It is also interesting to investigate the interplay between the direct and indirect DM searches and low energy precision observables, such as anomalous magnetic moment of leptons and the lepton flavor violating decays.
\section*{Acknowledgement}

This work is supported in part by the National Science Foundation of China (No. 11075139 and No.10705024) and
National Basic Research Program of China (No. 2010CB833000). G.Z is also supported in part by
the Fundamental Research Funds for the Central Universities.


\begin{thebibliography}{99}
\bibitem{Hooper:2010mq}
  D.~Hooper and L.~Goodenough,
  arXiv:1010.2752 [hep-ph].

\bibitem{Finkbeiner:2003im}
  D.~P.~Finkbeiner,
  Astrophys.\ J.\  {\bf 614} (2004) 186
  [arXiv:astro-ph/0311547].

\bibitem{Dobler:2007wv}
  G.~Dobler and D.~P.~Finkbeiner,
  Astrophys.\ J.\  {\bf 680} (2008) 1222
  [arXiv:0712.1038 [astro-ph]].

\bibitem{Hooper:2010im}
  D.~Hooper and T.~Linden,
  arXiv:1011.4520 [astro-ph.HE].

\bibitem{Crocker:2010qn}
  R.~M.~Crocker, D.~I.~Jones, F.~Aharonian, C.~J.~Law, F.~Melia, T.~Oka and J.~Ott,
  arXiv:1011.0206 [astro-ph.GA].



\bibitem{Aalseth:2010vx}
  C.~E.~Aalseth {\it et al.}  [CoGeNT collaboration],
  arXiv:1002.4703 [astro-ph.CO].

\bibitem{Bernabei:2010mq}
  R.~Bernabei {\it et al.},
  Eur.\ Phys.\ J.\  C {\bf 67} (2010) 39
  [arXiv:1002.1028 [astro-ph.GA]].

\bibitem{Kopp:2009qt}
  J.~Kopp, T.~Schwetz and J.~Zupan,
  JCAP {\bf 1002} (2010) 014
  [arXiv:0912.4264 [hep-ph]].

\bibitem{Bottino:2009km}
  A.~Bottino, F.~Donato, N.~Fornengo and S.~Scopel,
  Phys.\ Rev.\  D {\bf 81} (2010) 107302
  [arXiv:0912.4025 [hep-ph]].

\bibitem{Fitzpatrick:2010em}
  A.~L.~Fitzpatrick, D.~Hooper and K.~M.~Zurek,
  Phys.\ Rev.\  D {\bf 81} (2010) 115005
  [arXiv:1003.0014 [hep-ph]].

\bibitem{Chang:2010yk}
  S.~Chang, J.~Liu, A.~Pierce, N.~Weiner and I.~Yavin,
  JCAP {\bf 1008} (2010) 018
  [arXiv:1004.0697 [hep-ph]].

\bibitem{Foot:2010rj}
  R.~Foot,
  Phys.\ Lett.\  B {\bf 692} (2010) 65
  [arXiv:1004.1424 [hep-ph]].


\bibitem{Aprile:2010um}
  E.~Aprile {\it et al.}  [XENON100 Collaboration],
  Phys.\ Rev.\ Lett.\  {\bf 105} (2010) 131302
  [arXiv:1005.0380 [astro-ph.CO]].

\bibitem{Ahmed:2010wy}
  Z.~Ahmed {\it et al.}  [CDMS-II Collaboration],
  arXiv:1011.2482 [astro-ph.CO].

\bibitem{Ahmed:2009zw}
  Z.~Ahmed {\it et al.}  [The CDMS-II Collaboration],
  Science {\bf 327} (2010) 1619
  [arXiv:0912.3592 [astro-ph.CO]].

\bibitem{Lisanti:2010qx}
  M.~Lisanti, L.~E.~Strigari, J.~G.~Wacker and R.~H.~Wechsler,
  arXiv:1010.4300 [astro-ph.CO].

\bibitem{Savage:2010tg}
  C.~Savage, G.~Gelmini, P.~Gondolo and K.~Freese,
  arXiv:1006.0972 [astro-ph.CO].

\bibitem{Collar:2010nx}
  J.~I.~Collar,
  arXiv:1006.2031 [astro-ph.CO].



\bibitem{Collar:2010ht}
  J.~I.~Collar,
  arXiv:1010.5187 [astro-ph.IM].

\bibitem{Hooper:2010uy}
  D.~Hooper, J.~I.~Collar, J.~Hall, D.~McKinsey and C.~M.~Kelso,
  Phys.\ Rev.\  D {\bf 82} (2010) 123509
  [arXiv:1007.1005 [hep-ph]].

\bibitem{Bottino:2003iu}
  A.~Bottino, F.~Donato, N.~Fornengo and S.~Scopel,
  Phys.\ Rev.\  D {\bf 68} (2003) 043506
  [arXiv:hep-ph/0304080].

\bibitem{Bottino:2003cz}
  A.~Bottino, F.~Donato, N.~Fornengo and S.~Scopel,
  Phys.\ Rev.\  D {\bf 69} (2004) 037302
  [arXiv:hep-ph/0307303].



\bibitem{Fornengo:2010mk}
  N.~Fornengo, S.~Scopel and A.~Bottino,
  Phys.\ Rev.\  D {\bf 83} (2011) 015001
  [arXiv:1011.4743 [hep-ph]].

\bibitem{Scopel:2011qt}
  S.~Scopel, S.~Choi, N.~Fornengo and A.~Bottino,
  arXiv:1102.4033 [hep-ph].

\bibitem{Feng:2008ya}
  J.~L.~Feng and J.~Kumar,
  Phys.\ Rev.\ Lett.\  {\bf 101} (2008) 231301
  [arXiv:0803.4196 [hep-ph]].

\bibitem{Feng:2008mu}
  J.~L.~Feng, H.~Tu and H.~B.~Yu,
  JCAP {\bf 0810} (2008) 043
  [arXiv:0808.2318 [hep-ph]].

\bibitem{Khlopov:2010pq}
  M.~Y.~Khlopov, A.~G.~Mayorov and E.~Y.~Soldatov,
  Int.\ J.\ Mod.\ Phys.\  D {\bf 19} (2010) 1385
  [arXiv:1003.1144 [astro-ph.CO]].

\bibitem{Feng:2008dz}
  J.~L.~Feng, J.~Kumar and L.~E.~Strigari,
  Phys.\ Lett.\  B {\bf 670} (2008) 37
  [arXiv:0806.3746 [hep-ph]].

\bibitem{Barger:2010ng}
  V.~Barger, J.~Kumar, D.~Marfatia and E.~M.~Sessolo,
  Phys.\ Rev.\  D {\bf 81} (2010) 115010
  [arXiv:1004.4573 [hep-ph]].

\bibitem{McKeen:2009rm}
  D.~McKeen,
  Phys.\ Rev.\  D {\bf 79} (2009) 114001
  [arXiv:0903.4982 [hep-ph]].

\bibitem{McKeen:2009ny}
  D.~McKeen,
  arXiv:0912.1076 [hep-ph].

\bibitem{Buckley:2010ve}
  M.~R.~Buckley, D.~Hooper and T.~M.~P.~Tait,
  arXiv:1011.1499 [hep-ph].

\bibitem{Alwall:2010jc}
  J.~Alwall, J.~L.~Feng, J.~Kumar and S.~Su,
  Phys.\ Rev.\  D {\bf 81} (2010) 114027
  [arXiv:1002.3366 [hep-ph]].

\bibitem{LEPSUSY}
LEPSUSYWG, ALEPH, DELPHI, L3 and OPAL experiments, note LEPSUSYWG/04-01
(http://lepsusy.web.cern.ch/lepsusy/).

\bibitem{Bottino:2001dj}
  A.~Bottino, F.~Donato, N.~Fornengo and S.~Scopel,
  Astropart.\ Phys.\  {\bf 18} (2002) 205
  [arXiv:hep-ph/0111229].

\bibitem{Ellis:2005mb}
  J.~R.~Ellis, K.~A.~Olive, Y.~Santoso and V.~C.~Spanos,
  Phys.\ Rev.\  D {\bf 71} (2005) 095007
  [arXiv:hep-ph/0502001].

\bibitem{Nakamura:2010zzi}
  K.~Nakamura {\it et al.}  [Particle Data Group],
  J.\ Phys.\ G {\bf 37} (2010) 075021.

\bibitem{Aoyama:2007mn}
  T.~Aoyama, M.~Hayakawa, T.~Kinoshita and M.~Nio,
  Phys.\ Rev.\  D {\bf 77} (2008) 053012
  [arXiv:0712.2607 [hep-ph]].





\end{thebibliography}
\end{document}